\documentclass[11pt,a4paper]{amsart}

\usepackage[latin1]{inputenc}
\usepackage[english]{babel}
\usepackage{amsmath}

\usepackage{amssymb, mathabx}
\usepackage[numbers]{natbib}
\usepackage{graphicx}
\usepackage{braket}
\usepackage[pdftex,plainpages=false,colorlinks,hyperindex,bookmarksopen,linkcolor=red,citecolor=blue,urlcolor=blue]{hyperref}
\usepackage{caption}
\usepackage{subcaption}

\usepackage{mathrsfs}

\usepackage{epstopdf}

\usepackage{braket}

\bibpunct{[}{]}{;}{n}{,}{,}
\usepackage[hmargin=3cm,vmargin={3.5cm,4cm}]{geometry}

\theoremstyle{theorem}

\theoremstyle{definition}                                 

\theoremstyle{definition}                           

\theoremstyle{remark}                             

\usepackage{color}

\usepackage{mathtools,slashed}

\newcommand{\be}{\begin{eqnarray}}
\newcommand{\ee}{\end{eqnarray}}

\newcommand{\R}{\mathbb{R}}  

\newcommand{\wt}[1]{\widetilde{#1}}
\newcommand{\wh}[1]{\widehat{#1}}

\numberwithin{equation}{section}

\allowdisplaybreaks

\begin{document}
\title{Storage and Dissipation of Energy\\ in Prabhakar Viscoelasticity}
	
	   \author{Ivano Colombaro$^1$}
		\address{${}^1$ Department of Information and Communication Technologies, Universitat Pompeu Fabra and INFN.
		C/Roc Boronat 138, Barcelona, SPAIN.}
		\email{ivano.colombaro@upf.edu}
	
	    \author{Andrea Giusti$^2$}
		\address{${}^2$ Department of Physics $\&$ Astronomy, University of 	
    	    Bologna and INFN. Via Irnerio 46, Bologna, ITALY and 
	    	 Arnold Sommerfeld Center, Ludwig-Maximilians-Universit\"at, 
	    	 Theresienstra{\ss}e~37, 80333 M\"unchen, GERMANY.   
    	    }
		\email{andrea.giusti@bo.infn.it}
	
    \author{Silvia Vitali$^3$}
    	    \address{${}^3$ Department of Physics $\&$ Astronomy, University of 	
    	    Bologna, Via Irnerio 46, Bologna, ITALY.}
			\email{silvia.vitali4@unibo.it}
 

	\thanks{
			In: \textbf{Mathematics~(2018),  6(2), 15
}, \textbf{DOI}: \href{http://www.mdpi.com/2227-7390/6/2/15/}{10.3390/math6020015}
			}	
	
    \date  {\today}

\begin{abstract}
In this paper, after a brief review of the physical notion of quality factor in viscoelasticity, we present a complete discussion of the attenuation processes emerging in the Maxwell--Prabhakar model, recently developed by Giusti and Colombaro. Then, taking profit of some illuminating plots, we discuss some potential connections between the presented model and the modern mathematical modelling of seismic processes.  
\end{abstract}

    \maketitle

\section{Introduction} \label{sec-intro}
	The linear theory of viscoelasticity, despite its apparent simplicity, keeps having a striking impact in geophysics, theoretical mechanics and biophysics; see, for example,  \cite{IC-AG-FM-TWLV, IC-AG-FM-ZAMP, IC-AG-FM-Bessel, Garra-Mainardi-Spada, AG-FCAA-2017, AG-FM_MECC16, Mainardi-1997}. Besides, fractional calculus \cite{Giusti-Arxiv, Mainardi-Gorenflo-1997, Mainardi-Book} has proven itself to be one of the fundamental languages for describing processes involving memory effects, like the ones that are typically featured by viscoelastic systems. Concerning the latter, it is also worth remarking on the pivotal role of the notion of complete monotonicity, which was first (implicitly) hinted at by Gross \cite{Gross53} in 1953, and then brought to light by Molinari \cite{Molinari}, in 1973. These~seminal studies were then followed by many other authors; see, for example,   \cite{Mainardi-Book, Mainardi-Turchetti, Hanyga-1}.

	A simple generalization of the well-known fractional Maxwell model of linear viscoelasticity was first introduced by Giusti and Colombaro in \cite{GC-CNSNS}. In this paper the authors provide an extension of the classical model by replacing the Caputo fractional derivative with the Prabhakar one in the constitutive equation. Concretely, if we denote with $\sigma , \varepsilon$, the stress and the strain for a given system, respectively, and we further assume that these functions are both causal such that $\sigma , \varepsilon \in AC^1 \left(0, \, + \infty \right)$, then the constitutive equation of the Maxwell--Prabhakar model \cite{GC-CNSNS} reads
	\be \label{P-maxwell}
	\sigma (t) + a \, ^C \textbf{D} ^\gamma _{\alpha , \beta , \Omega} \, \sigma (t) = 
	b \, ^C \textbf{D} ^\gamma _{\alpha , \beta , \Omega} \, \varepsilon (t) \, ,
	\ee
	where $a$ and $b$ are two suitable real constants and provided that $\alpha, \beta, \gamma, \Omega \in \R$, $\alpha > 0$ and $0 < \beta < 1$.
	 
	Here, $^C \textbf{D} ^\gamma _{\alpha , \beta , \Omega}$ represents the regularized Prabhakar derivative \cite{D'Ovidio-Polito}, which is defined by
	\be \label{P-derivative}
	^C _a \textbf{D} ^\gamma _{\alpha , \beta , \Omega} \, f (t) = 
	  {}_a \textbf{E} ^{-\gamma} _{\alpha , m - \beta , \Omega} \, f^{(m)} (t),
	\ee
	where
	\be \label{P-integral} 
	_a \textbf{E} ^\rho _{\mu , \nu , \lambda} \, f (t) = \int _a ^t (t - \tau ) ^{\nu - 1}  E ^\rho _{\mu , \nu} \left[ \lambda \, (t - \tau)^\mu \right] \, 
	f(\tau) \, d \tau \notag
	\, 
	\ee
	denotes the Prabhakar fractional integral \cite{GGPT} and $E ^\rho _{\mu , \nu} (t)$ represents the Prabhakar function \cite{GnG, FM-ML, Pippo, PBK}. Furthermore, it is important to stress that the function $t^{\nu - 1} \, E ^\rho _{\mu , \nu} (-t^\mu)$, for $t > 0$, is locally integrable and completely monotone provided that $0 < \mu \leq 1$ and $0 < \mu \, \rho \leq \nu \leq 1$; see, for example,   \cite{Capelas, MG}.
	
	It is important to remark that the Prabhakar fractional calculus has been attracting much attention in the mathematical community \cite{GnG, GGPT, Garrappa, RG-FM-GM, GC-CNSNS, MG, Sandev}, particularly because of its connection with the theoretical description of the Havriliak--Negami model \cite{GnG, Garrappa, Hanyga-2, Hanyga-3}. Moreover, this growing interest in Prabhakar's calculus is also reflected by the increasing literature on the recently proposed Maxwell--Prabhakar model, which was also kindly referred to as the Giusti--Colombaro model in \cite{ScientificReports}.
	
	In this paper we wish to analyze the important phenomena of storage and dissipation of energy in linear viscoelastic media, with particular regard for the class of models emerging from the constitutive equation in Equation~\eqref{P-maxwell}. In viscoelasticity, as well as in electrical engineering, the process of dissipation of energy is usually accounted for in terms of a dimensionless parameter, called the quality factor, that is roughly defined as the ratio of the peak of energy stored in the system under a cycle of forced harmonic oscillation to the total rate of change of the energy, per cycle, by damping processes. Therefore, the aim of this paper is to compute and discuss the quality factor for the model defined in Equation~\eqref{P-maxwell}.

\section{Storage and Dissipation of Energy in Linear Viscoelasticity} \label{sec-intro-Q}
	In this section we wish to review the general theory, concerning the theoretical foundations, that leads to the definition of quality factor for a viscoelastic system. In order to do so, we will mimic the arguments presented in \cite{Book, Mainardi-Book}, unifying these formulations according to the notations employed in this paper.
	
	Let us consider a quiescent viscoelastic body for $t < 0$. Then, under the hypothesis of sufficiently well-behaved causal histories, its constitutive equation in the creep representation reads
	\be \label{eq-gen-constitutive}
	\varepsilon (t) = \int _0 ^t J (t - \tau) \,  {\text{d}}\sigma (\tau) = \sigma (0+) \, J(t) + \int_0 ^t J (t - \tau) \, \dot{\sigma} (\tau) \, {\text{d}}\tau \, ,
	\ee
where ${\text{d}}\sigma (\tau)$ represents the Riemann--Stieltjes measure and $J(t)$ is the so-called creep compliance of the system, that in the Laplace domain is given by
\be 
\wt{\varepsilon} (s) = s \, \wt{J} (s) \, \wt{\sigma} (s) \, .
\ee 

In order to consider the harmonic behavior of a linear viscoelastic material, we should assume that a sufficient amount of time has elapsed since the original perturbation so that the effect of initial conditions could be considered negligible. So, let us consider some harmonic excitation of the material, which can be described in terms of the complex exponential representation, that is, 
\be \label{eq-sigma-harm}
\sigma (t \, ; \, \omega) = \chi \, \exp \left(i \, \omega \, t \right) \, , \,\,\, \omega > 0 \, , \,\, - \infty < t < \infty \, , \,\, \chi \in \mathbb{C} \, .
\ee
Clearly, a similar argument can be presented in terms of the relaxation representation, however we will only focus on the creep one for sake of brevity.

	If we now plug \eqref{eq-sigma-harm} into \eqref{eq-gen-constitutive} we get
	\be 
	\varepsilon (t \, ; \, \omega) = i \, \omega \, \wh{J} (\omega) \, (\chi \, \exp \left(i \, \omega \, t \right)) = i \, \omega \, \wh{J} (\omega) \, \sigma (t \, ; \, \omega) \, ,
	\ee
where $\wh{J} (\omega)$ stands for the Fourier transform of $J (t)$ that, the latter being a causal function, ultimately~reads
$$ \wh{J} (\omega) = \int _0 ^\infty \exp (- i \, \omega \, t ) \, J (t) \, {\text{d}} t \, . $$

Moreover, if we denote $J ^\star (\omega) = i \, \omega \, \wh{J} (\omega)$, as in \cite{Mainardi-Book}, the constitutive equation, in the creep representation, for a viscoelastic body subject to a harmonic stress excitation reduces to
\be \label{eq-cost-harm}
\varepsilon (t \, ; \, \omega) =  J ^\star (\omega) \, \sigma (t \, ; \, \omega) \, .
\ee

	The time rate of change of energy in the system is then given by
	\be \label{eq-work}
	\dot{\mathcal{W}} = \sigma _{\rm R} (t \, ; \, \omega) \, \dot{\varepsilon} _{\rm R} (t \, ; \, \omega) \, ,
	\ee
where the subscript ${\rm R}$ indicates that we are considering the real part of the corresponding function.

	Now, one can easily solve \eqref{eq-cost-harm} for $\sigma (t \, ; \, \omega)$, then taking the real part of the resulting equation gives
	\be \label{eq-sigma-r}
	\sigma _{\rm R} (t \, ; \, \omega) = \frac{\varepsilon _{\rm R} (t \, ; \, \omega) \, J^\star _{\rm R} (\omega) - \varepsilon _{\rm I} (t \, ; \, \omega) \, J^\star _{\rm I} (\omega)}{|J^\star (\omega)|^2} \, ,
	\ee
where we denoted $\sigma \equiv \sigma _{\rm R} + i \, \sigma _{\rm I}$, $\varepsilon \equiv \varepsilon _{\rm R} + i \, \varepsilon _{\rm I}$ and $J ^\star (\omega) \equiv J ^\star _{\rm R} (\omega) - i \, J ^\star _{\rm I} (\omega)$ for future convenience. Besides, from \eqref{eq-sigma-harm} and \eqref{eq-cost-harm} it is also easy to see that
\be \label{eq-dot-strain}
\varepsilon _{\rm I} (t \, ; \, \omega)  = - \frac{\dot{\varepsilon} _{\rm R} (t \, ; \, \omega)}{\omega}  \, .
\ee

	Hence, if we plug \eqref{eq-sigma-r} into \eqref{eq-work} and recall the result in \eqref{eq-dot-strain}, then after some simple manipulations $\dot{\mathcal{W}}$ reduces to
	\be \label{eq-work-explicit}
	\dot{\mathcal{W}} = \frac{\partial}{\partial t} \left[ \frac{1}{2} \, \frac{J ^\star _{\rm R} (\omega)}{|J^\star (\omega)|^2} \, \varepsilon _{\rm R}  ^2 \right] + \frac{1}{\omega}\, \frac{J ^\star _{\rm I} (\omega)}{|J^\star (\omega)|^2} \,  \dot{\varepsilon} _{\rm R}  ^2 \, , 
	\ee
where we omitted the explicit dependence on $t$ and $\omega$ in the strain for sake of clarity.

	Thus, it is easy to see that the total rate of change of energy over one cycle is accounted for by the integral over the cycle of the second term on the right-hand side of \eqref{eq-work-explicit}, namely,
\be 
\frac{\Delta \mathcal{E}}{\mbox{Cycle}} = \int _t ^{t + T}  \dot{\mathcal{W}} (\tau) \, {\text{d}} \tau = \pi \, \frac{J ^\star _{\rm I} (\omega)}{|J^\star (\omega)|^2} \, |\chi| ^2\, ,
\ee
with $T = 2 \pi / \omega$ the period of the cycle.

	Due to the second law of thermodynamics, which requires that the total amount of energy dissipated increases with time, one can further infer that $J ^\star _{\rm I} (\omega) \geq 0$.

	However, despite defining a boundary term, the first piece of the right-hand side of \eqref{eq-work-explicit} carries a very important physical meaning. Indeed, it tells us that the peak energy stored during a cycle is given~by
	\be 
	\mathcal{P} _{\rm max} = \frac{1}{2} \, \frac{J ^\star _{\rm R} (\omega)}{|J^\star (\omega)|^2} \, |\chi| ^2\, .
	\ee

One can now define the specific attenuation factor, or quality factor ($Q$-factor), as a normalized non-dimensional quantity defined by    
\be 
Q^{-1} \equiv \frac{1}{2 \, \pi} \frac{\Delta \mathcal{E} / \mbox{Cycle}}{\mathcal{P} _{\rm max}} \, .
\ee
	
	Then, taking profit of the previous discussion it is easy to see that
	\be 
	Q^{-1} = \frac{1}{2 \, \pi} \frac{\Delta \mathcal{E} / \mbox{Cycle}}{\mathcal{P} _{\rm max}} = \frac{J ^\star _{\rm I} (\omega)}{J ^\star _{\rm R} (\omega)} = - \frac{\Im \left\{ i \, \omega \, \wh{J} (\omega) \right\}}{\Re \left\{ i \, \omega \, \wh{J} (\omega) \right\}} \, ,
	\ee
recalling that $J ^\star _{\rm I} (\omega) = - \Im \left\{ J ^\star (\omega) \right\}$.
	
If we combine the fact that the Fourier transform is equivalent to evaluating the bilateral Laplace transform with imaginary argument $s = i \, \omega$, together with the assumption for which $J(t)$ is a causal function, one can conclude that $\wh{J} (\omega) = \wt{J} (s) \, | _{s = i\, \omega}$. This argument ultimately leads us to a very useful expression for the $Q$-factor, namely,
\be 
Q^{-1} (\omega) = - \frac{\Im \left\{s \, \wt{J} (s) \, | _{s = i\, \omega} \right\}}{\Re \left\{ s \, \wt{J} (s) \, | _{s = i\, \omega} \right\}} \, ,
\ee
where we shall consider some positive real frequencies $\omega$.

\section{Quality Factor in Prabhakar-Like Viscoelasticity} \label{sec-Q-P}
	Let us now compute the $Q$-factor for the Maxwell--Prabhakar model. Recalling that the Laplace transform of the Prabhakar integral kernel is given by
	\be \label{eq-LT-gml}
	\mathcal{L} \left\{ t^{\beta -1} \, E^{\gamma} _{\alpha , \, \beta} ( \lambda \, t^\alpha ) \right\} 
	= 
	s^{- \beta} \, \left( 1 - \lambda \, s^{-\alpha} \right) ^{-\gamma}
	\, \, ,
	\ee
	where $t \in \R$, $\alpha, \beta, \gamma, \lambda \in \mathbb{C}$ and $\texttt{Re}(\beta) > 0$, then it is easy to see that the creep compliance, in the Laplace domain, for a system described in terms of Equation~\eqref{P-integral} is therefore given by
	\be \label{eq-sJs}
	s \, \wt{J} (s) = \frac{a}{b} + \frac{1}{b \, s^\beta \, \left( 1 - \Omega \, s^{-\alpha} \right) ^\gamma} \, .
	\ee 
Then, if we apply the replacement $s = i \, \omega$, the latter turns into
\be 
J^\star (\omega) = i \, \omega \, \wh{J} (\omega) = \frac{a}{b} + \frac{1}{b \, (i \, \omega)^\beta \, \left[ 1 - \Omega \, (i \, \omega)^{-\alpha} \right] ^\gamma} \, .
\ee

	Let us define an auxiliary variable $z(\omega) =  1 - \Omega \, (i \, \omega)^{-\alpha}$. Then, considering $\omega \in \R^+$, one can easily rewrite $s = i \, \omega = \omega \, \exp (i \, \pi /2)$, where $|s| = \omega$, that allows us to recast this new variable in the exponential representation, that is, 
\be
z (\omega) = |z (\omega)| \, \exp \left( i \, \theta _z (\omega) \right) \, ,
\ee
with
\be 
|z (\omega)| ^2 = 1 + \frac{\Omega ^2}{\omega ^{2\alpha}} - \frac{2 \, \Omega}{\omega ^{\alpha}} \, \cos \left( \frac{\alpha \, \pi}{2}\right) \, ,
\ee
\be 
\theta _z (\omega) = \arctan \left[ \frac{\Omega \, \sin \left(\alpha \, \pi / 2 \right)}{\omega^\alpha - \Omega \, \cos \left(\alpha \, \pi / 2 \right)} \right] \, .
\ee

	Then, plugging $z (\omega)$ into Equation~\eqref{eq-sJs}, one can easily infer that
\be 
J^\star (\omega) = \frac{a}{b} + \frac{1}{b \, \omega ^\beta \, |z(\omega)|^\gamma} \, \exp \left[- i \, \left(\frac{\beta \, \pi}{2} + \gamma \, \theta _z (\omega) \right) \right] \, ,
\ee	
from which we can conclude that
\be 
\Re \left\{ J^\star (\omega) \right\} = \frac{a}{b} + \frac{1}{b \, \omega ^\beta \, |z(\omega)|^\gamma} \, \cos \left(\frac{\beta \, \pi}{2} + \gamma \, \theta _z (\omega) \right) \, ,
\ee
and
\be 
\Im \left\{ J^\star (\omega) \right\} = - \frac{1}{b \, \omega ^\beta \, |z(\omega)|^\gamma} \, \sin \left(\frac{\beta \, \pi}{2} + \gamma \, \theta _z (\omega) \right) \, .
\ee
Hence, the quality factor for a Maxwell--Prabhakar viscoelastic body is given by 
\be \label{eq-Q-P}
Q^{-1} (\omega) = \frac{\sin \left(\frac{\beta \, \pi}{2} + \gamma \, \theta _z (\omega) \right)}{a \, \omega ^\beta \, |z(\omega)|^\gamma + \cos \left(\frac{\beta \, \pi}{2} + \gamma \, \theta _z (\omega) \right)} \, .
\ee

\section{Quality Factor for Some Specific Realizations of the Maxwell--Prabhakar Model} \label{sec-examples}
	In this section we discuss the quality factor for different choices of the parameters of the discussed model. Specifically, we will focus our discussion on four cases corresponding to two well-known classical viscoelastic models and the viscoelastic analogue of the Havriliak--Negami model for dielectric~relaxation.
	
	\subsection{Fractional Maxwell Model}
	As argued in \cite{GC-CNSNS}, it is easy to see that Equation~\eqref{P-maxwell} naturally reduces to the fractional Maxwell model, namely,
	\be 
	\sigma (t) + a \, ^C \textbf{D} ^\nu \, \sigma (t) = 
	b \, ^C \textbf{D} ^\nu \, \varepsilon (t) \, ,
	\ee
where $^C \textbf{D} ^\nu$ represents the Caputo fractional derivative, provided that the parameters are chosen according to one of these two configurations:
	\begin{itemize}
	\item[(i)] $\gamma = 0$, $a > 0$, $b >0$, $\beta = \nu$, $\Omega \in \R$;
	\item[(ii)] $\gamma \in \R$, $a > 0$, $b >0$, $\beta = \nu$, $\Omega = 0$.
	\end{itemize} 

Here, it is trivial to infer that if $\gamma = 0$ then neither $|z(\omega)|$ nor $\theta _z (\omega)$ enter in the expression for the $Q$-factor, whereas if $\Omega = 0$ it is easy to see that $|z(\omega)| = 1$ and $\theta _z (\omega) = 0$. Hence, we get
\be 
Q^{-1} (\omega) = \frac{\sin \left(\nu \pi /2\right)}{a\, \omega ^\nu + \cos \left(\nu \pi /2\right)} \,.
\ee

Furthermore, it is also worth remarking that if we set $\nu = 1$ (ordinary limit), we explicitly recover the $Q$-factor for the (ordinary) Maxwell model, that is,  $Q^{-1} (\omega) = (a \, \omega)^{-1}$ (see, for example,   \cite{Mainardi-Book}).

\subsection{Fractional Voigt Model}	
	Again, following the analysis presented in \cite{GC-CNSNS}, one has that Equation~\eqref{P-maxwell} reduces to the fractional Voigt model, that is,
	\be 
	\sigma (t) = M \, \varepsilon (t) + B \, ^C \textbf{D} ^\nu \, \varepsilon (t) \, ,
	\ee
	 by setting $\gamma = 1$, $a=0$, $b=-B<0$, $\alpha = \beta = \nu $, $\Omega = - M/B < 0$.
	
	Now, if we plug this choice of parameters into Equation~\eqref{eq-Q-P}, we get
	\be
Q^{-1} (\omega) = \frac{\sin \left(\frac{\nu \, \pi}{2} + \theta _z (\omega) \right)}{\cos \left(\frac{\nu \, \pi}{2} + \theta _z (\omega) \right)} =
\tan \left(\frac{\nu \, \pi}{2} + \theta _z (\omega) \right) \, ,
	\ee
	with
	$$ \theta _z (\omega) = - \arctan \left[ \frac{M}{B} \frac{\sin \left(\nu \, \pi / 2 \right)}{\omega^\nu + (M/B) \, \cos \left(\nu \, \pi / 2 \right)} \right] \, . $$
	
The latter, in the limit for $\nu = 1$, reduces to
	\be
Q^{-1} (\omega) =
\tan \left( \frac{\pi}{2} + \theta _z (\omega) \right) = -\frac{1}{\tan \theta _z (\omega)} = \frac{B\, \omega}{M} \,.
	\ee
	which corresponds to the quality factor for the (ordinary) Voigt model (see, for example,   \cite{Mainardi-Book}).

\subsection{Havriliak--Negami Model}		
	The Havriliak--Negami relaxation is an empirical model which was first introduced in order to describe the dielectric relaxation of certain types of polymers \cite{Garrappa, RG-FM-GM, HN}. 
	
	Now, it is very well known that a viscoelastic system can usually be mapped onto a class of electrical ladder networks and vice versa; see, for example,   \cite{AG-FM-EPJP, Gross}.
	
	Following this line of thought, it is easy to see that the constitutive equation for the Havriliak--Negami viscoelastic model is given by \cite{GC-CNSNS},
	\be
\sigma (t) + a \, ^C \textbf{D} ^\gamma _{\alpha , \alpha \gamma , -\lambda} \, \sigma (t) = 
	b \, ^C \textbf{D} ^\gamma _{\alpha , \alpha \gamma , -\lambda} \, \varepsilon (t) \, ,
\ee
with $\beta = \alpha \gamma$, $\Omega = - \lambda$, with $\lambda > 0$, and $0 < \alpha , \, \gamma < 1$.
	
	Then, following the procedure presented in Section~\ref{sec-Q-P}, one finds that
	\be
	Q^{-1} (\omega) =   \frac{\sin \left[ \gamma \left(\frac{\alpha \, \pi}{2} + \theta _z (\omega) \right)\right]}{a \, \omega ^{\alpha\gamma} \, \left[ \omega ^{2\alpha} + \lambda ^2 + 2\lambda \omega ^\alpha \cos \left( \frac{\alpha \, \pi}{2} \right) \right]^{\gamma /2} + \cos \left[ \gamma \left(\frac{\alpha \, \pi}{2} + \theta _z (\omega) \right)\right]}
 \, , 
\ee
where
\be
\theta _z (\omega) = \arctan \left[ -\frac{\lambda \, \sin \left(\alpha \, \pi / 2 \right)}{\omega^\alpha + \lambda \, \cos \left(\alpha \, \pi / 2 \right)} \right] \, .
\ee

\section{Discussion and Conclusions}
	Attenuation effects represent one of the main fields of study in modern seismology, and consequently the specific attenuation factor (or $Q$-factor, for simplicity) embodies one of the key ingredients in geophysical sciences. Indeed, were it not for the damping capabilities of the soil, the energy of past earthquakes would still be resonating within the earth's interior.
	
\begin{figure}[h]
\centering
 \includegraphics[scale=0.34]{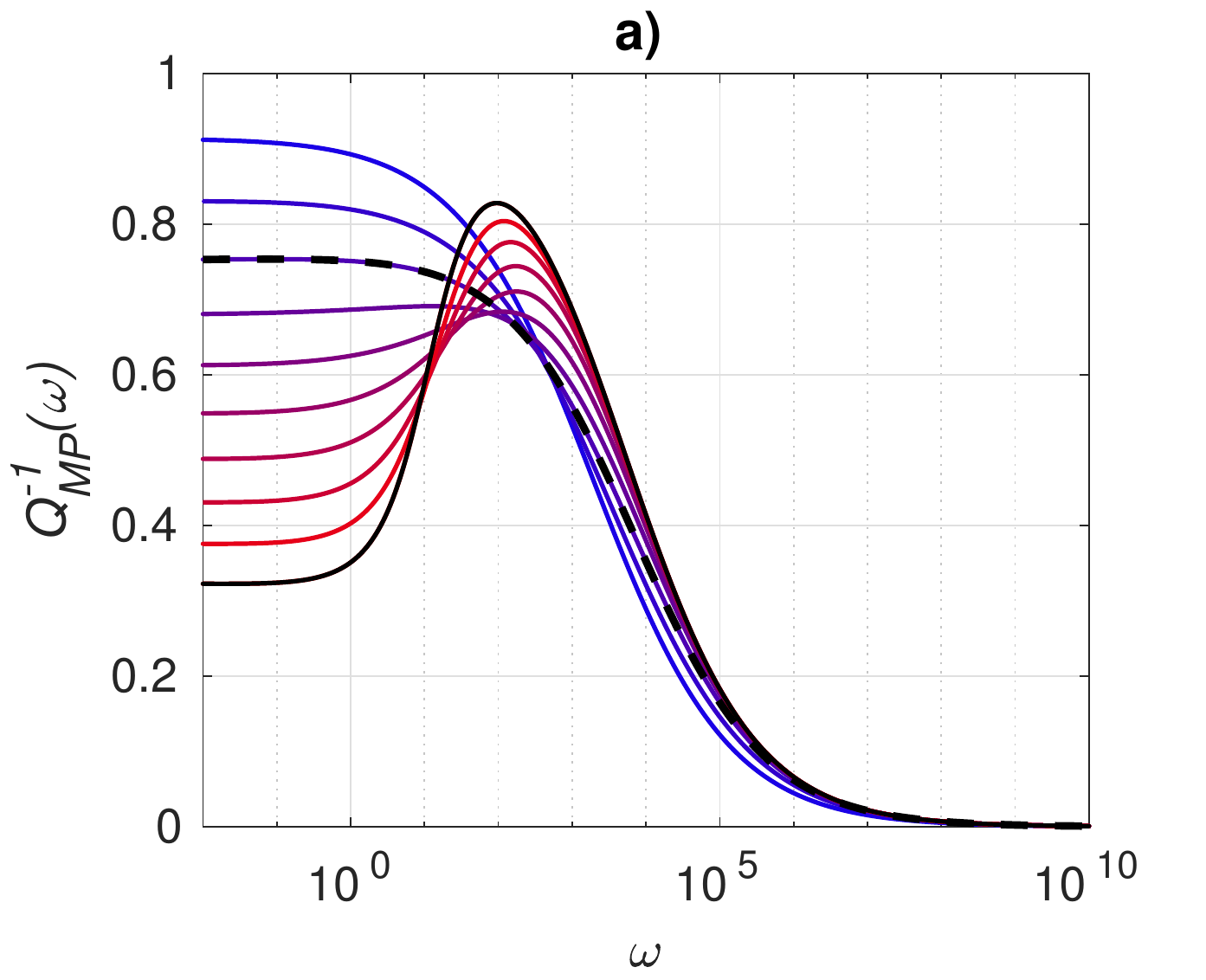} 
 \includegraphics[scale=0.34]{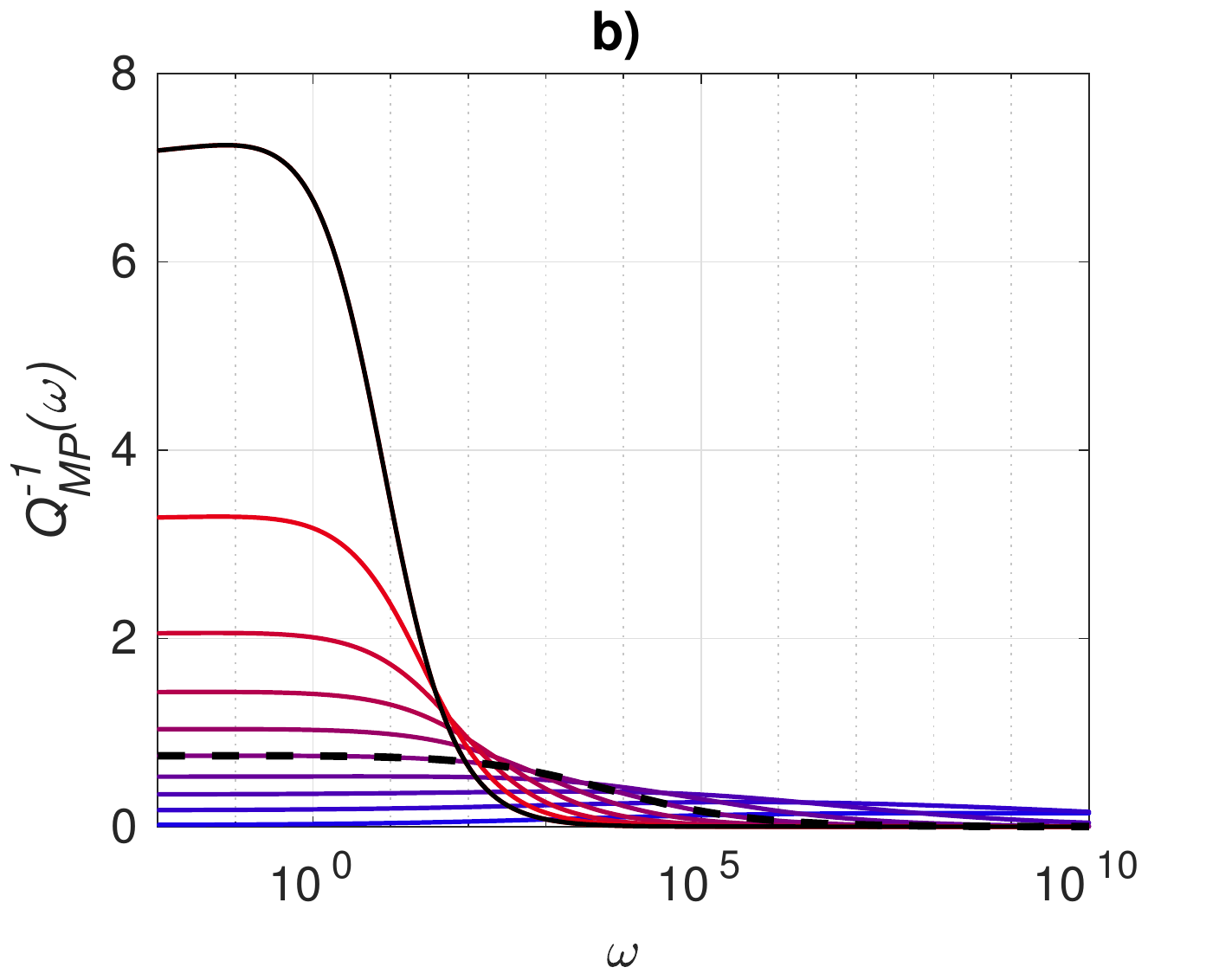} 
 \includegraphics[scale=0.34]{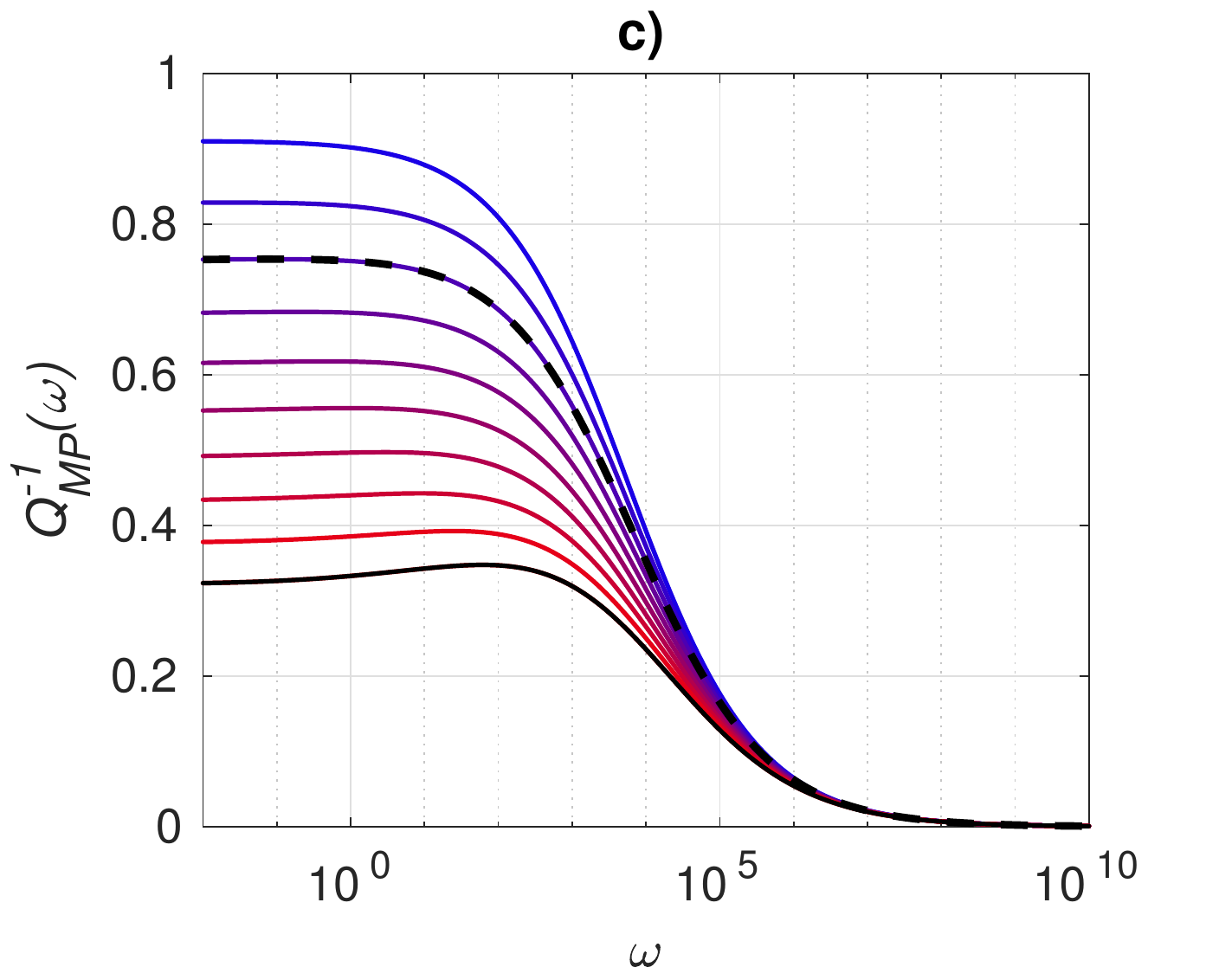} 
\caption[Maxwell--Prabhakar model]{{{{$Q$-factor of the Maxwell--Prabhakar model for $a=0.01 \, ,\Omega=-10$. Model dependence on} \mbox{$\alpha =0.1, 0.2, \ldots, 1$},  $\beta=0.5\,, \gamma=0.3$ (\textbf{a});
model dependence on \mbox{$\beta =0.1,0.2,\ldots,1$, $\alpha=0.3\,,\gamma=0.3$} (\textbf{b}); 
model dependence on $\gamma =0.1, 0.2, \ldots,1$,  $\beta=0.5\,,\alpha=0.3$ (\textbf{c}). Parameter value increasing from blue to red. In all panels the dashed line corresponds to $\alpha =0.3\,,\beta=0.5\,,\gamma=0.3$.}}}
\label{QMP}
\end{figure}

\begin{figure}[h]
\centering
 \includegraphics[scale=0.48]{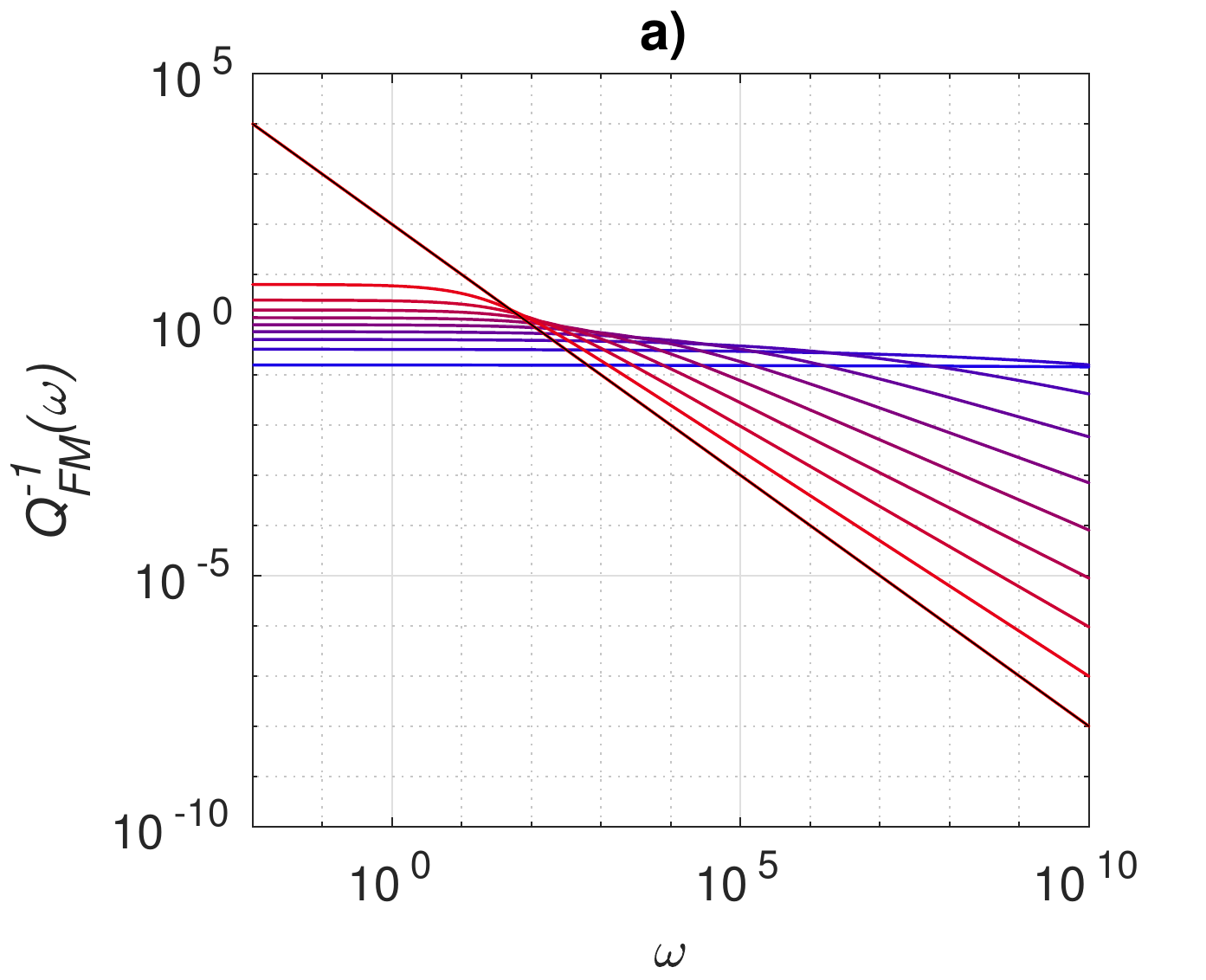} 
 \includegraphics[scale=0.48]{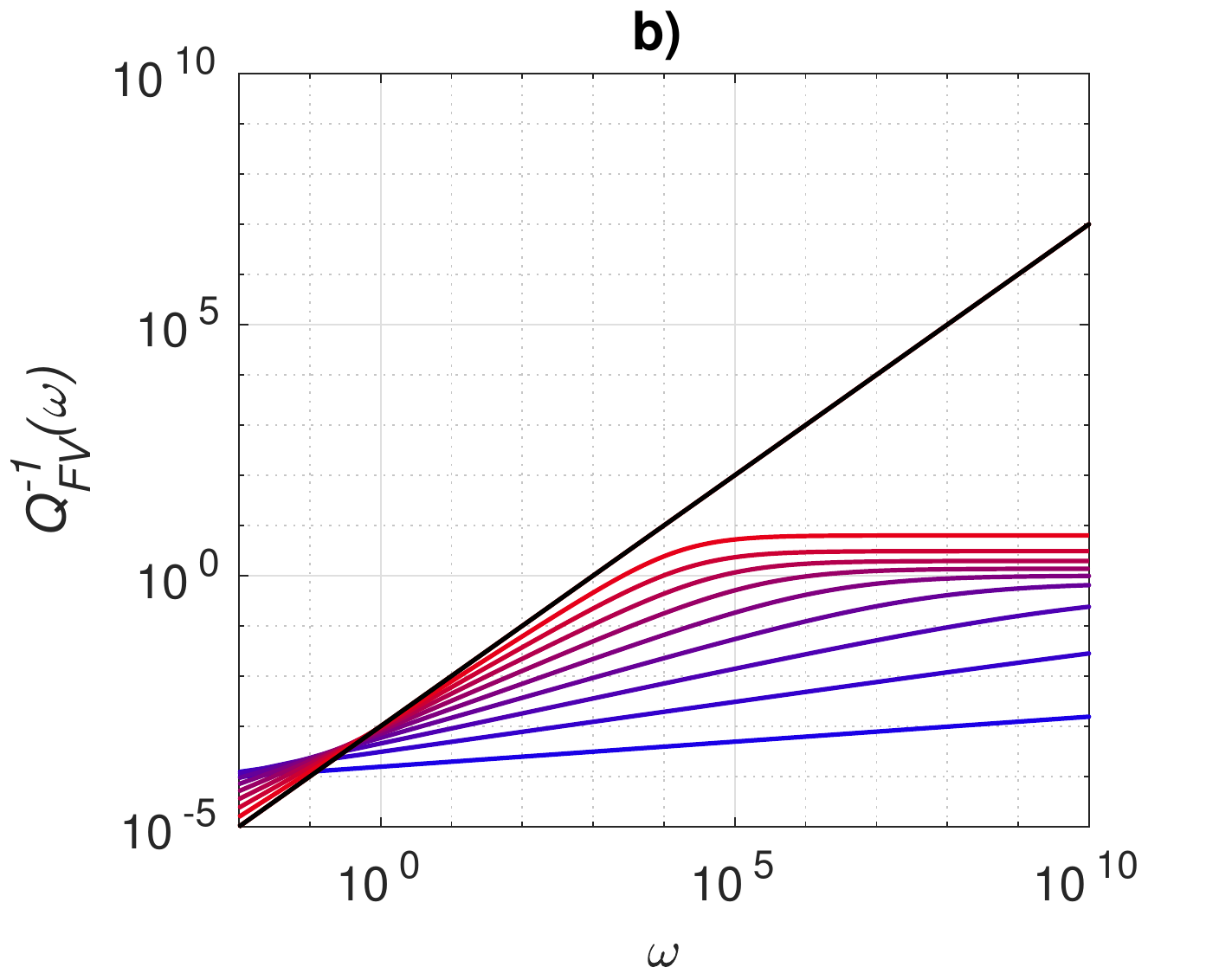} 

\caption[Fractional models]{$Q$-factor of fractional Maxwell model with $a=0.1$ (\textbf{a}), fractional Voigt model with $M=10^3, \, B=1$ (\textbf{b}). Parameter value increasing from blue to red.}
\label{FMVZ}   
\end{figure}

\begin{figure}[h]
\centering
 \includegraphics[scale=0.48]{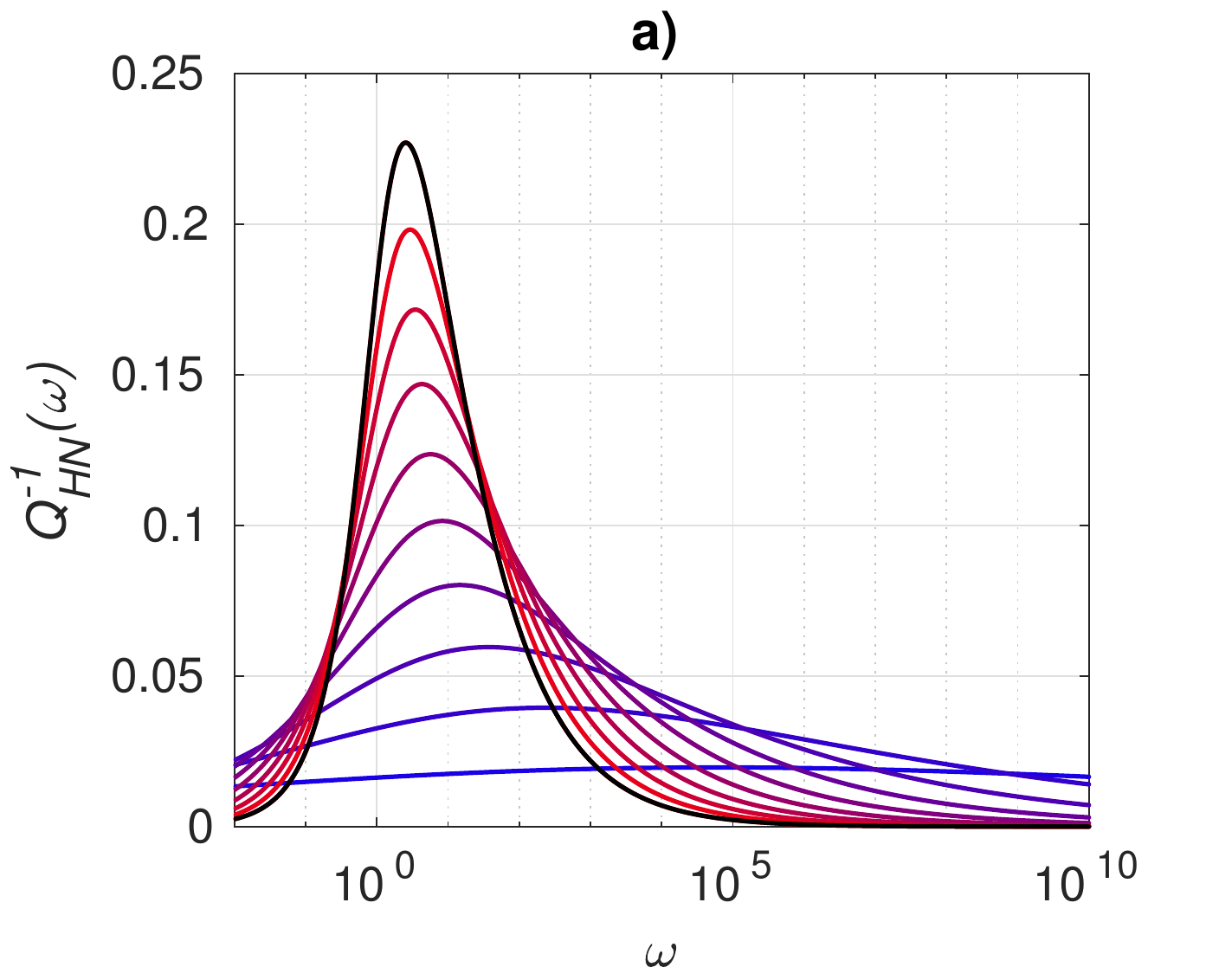}
 \includegraphics[scale=0.48]{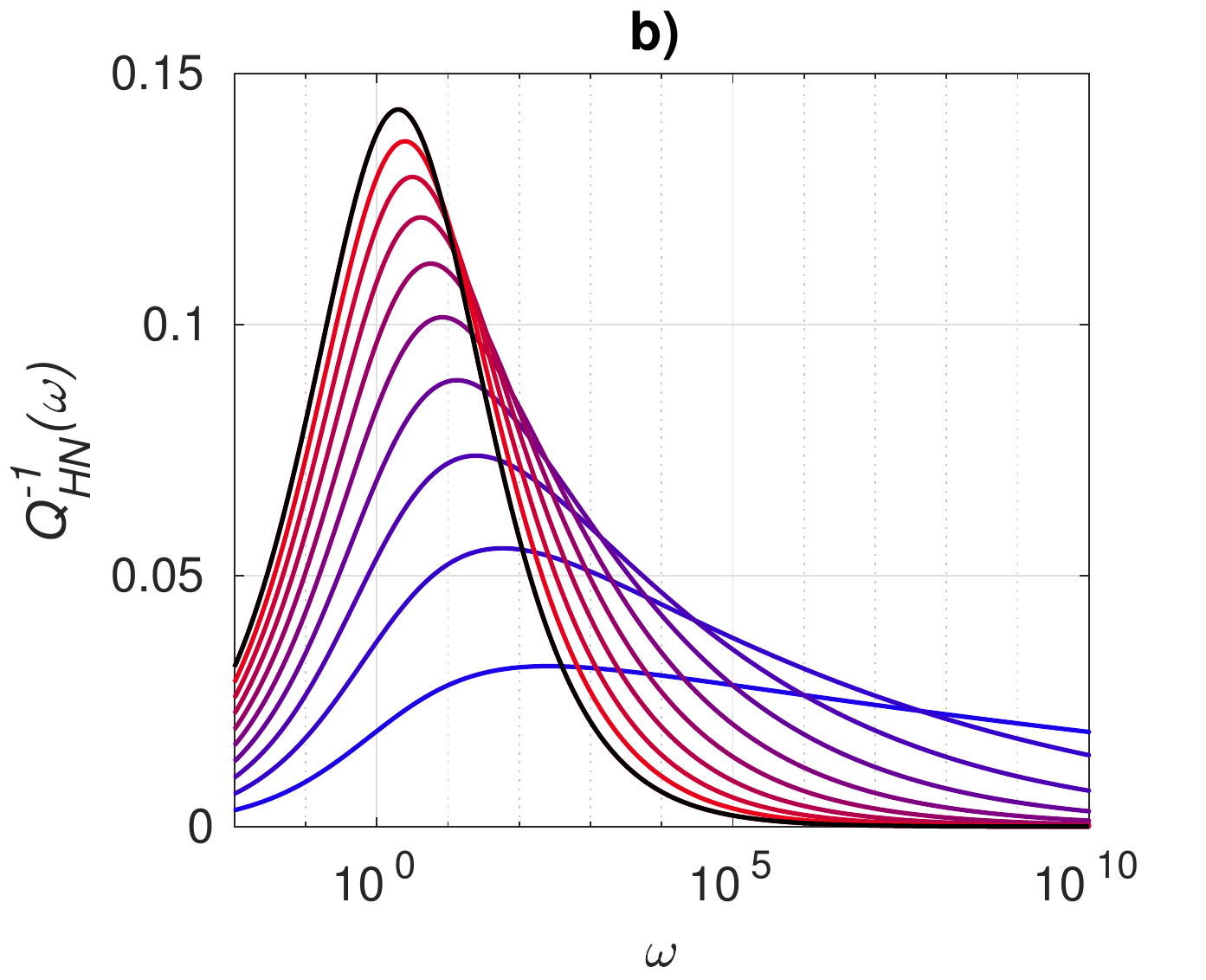} 
\caption[Havriliak-Negami model]{$Q$-factor of Havriliak--Negami model for dielectric relaxation with $a=1\,,\lambda=1$ and $\alpha =0.1,0.2,\ldots,1\,,\gamma=0.5$ (\textbf{a}),
or $a=1\,,\lambda=1$ and $\gamma=0.1,0.2,\ldots,1\,,\alpha =0.5$ (\textbf{b}). Parameter value increasing from blue to red. Decreasing $a$, the widths of the curves become wider, while increasing $\lambda$, the peak results are shifted to larger $\omega$ for lower $\alpha$. }
\label{QHN}
\end{figure}

	In this paper, after a thorough review of the physical definition and meaning of the quality factor $Q$, we have provided an analysis of the storage and dissipation of energy in viscoelastic material of the Maxwell--Prabhakar class, namely, the one featured by a constitutive relation given by Equation~\eqref{P-maxwell}. Specifically, in Section~\ref{sec-Q-P} we have computed the $Q$-factor for the general Maxwell--Prabhakar model, for which some interesting configurations of the parameters are shown in Figure~\ref{QMP}. In Section~~\ref{sec-examples} we have further provided some explicit realizations of the Maxwell--Prabhakar theory, namely the fractional Maxwell model, the fractional Voigt model and the viscoelastic equivalent of the Havriliak--Negami model for dielectric relaxation, shown in Figures~\ref{FMVZ} and \ref{QHN}.
	
	Let us pay particular attention to the cases displayed in Figure~\ref{QMP}. Indeed, as argued in \cite{Q}, there exists much experimental evidence supporting the theses for which the $Q$-factor of homogeneous materials is substantially independent of the frequency. In this respect, it is worth noting that the Maxwell--Prabhakar class shows a very slow varying (almost constant) behaviour of the quality factor for low frequencies, for certain choices of the parameters of the model (see Figure~\ref{QMP}). This is quite consistent with the results for the dumping of long-period teleseismic body waves and surface waves. Furthermore, for high frequencies, the model shows a power law behaviour, namely, $Q (\omega) \sim \omega ^\beta$ as $\omega \to \infty$, which appears to be consistent with the experimental results concerning the attenuation of the coda of high-frequency teleseismic waves in Earth's upper mantle \cite{Yale}. Furthermore, it is rather easy to prove that some very well-known constant-$Q$ models are nothing but some specific realizations of the model defined in Equation~\eqref{P-maxwell}. Indeed, for example, the renowned Kjartansson model \cite{Mainardi-e-amici, Kjartansson} can be obtained from \eqref{eq-sJs} by setting $\gamma = 0$, $\beta \equiv 2 \, \eta$ with $\eta \in (0, 1/2)$ and $a = 0$, which is nothing but the Scott--Blair model \cite{Mainardi-e-amici}. In view of the last few comments, we believe that the Maxwell--Prabhakar model of viscoelasicity can potentially provide some stimulating new insights into the mathematical modelling of seismic processes and therefore is worthy of further studies.
	
   \unskip	
	
  \unskip	

\newpage		

\section*{Acknowledgments}
The work of I.C. and A.G. has been carried out in the framework of the activities of the
National Group of Mathematical Physics (GNFM, INdAM). Moreover, the work of A.G. has been partially supported by GNFM/INdAM Young
Researchers Project 2017 ``Analysis of Complex Biological Systems''. Besides, the work of S.V. has been partially supported by the Interdepartmental Center ``Luigi Galvani'' for integrated studies of Bioinformatics, Biophysics and Biocomplexity of the University of Bologna.

\end{document}